\documentclass[12pt]{article}
\usepackage{amssymb}
\usepackage{amsmath}
\usepackage{apacite}
\usepackage{latexsym}
\usepackage{epsfig}
\usepackage{graphicx}
\usepackage{subfigure}
\usepackage{wasysym}
\usepackage{threeparttable}
\usepackage{natbib}
\usepackage{color}
\usepackage{epstopdf}
\usepackage{bm}
\usepackage{float}
\usepackage{todonotes}
\usepackage{verbatim}
\usepackage{arydshln}
\usepackage[symbol]{footmisc}
\usepackage{ulem}

\usepackage{hyperref}

\def\log{\hbox{log}}

\def\boxit#1{\vbox{\hrule\hbox{\vrule\kern6pt
          \vbox{\kern6pt#1\kern6pt}\kern6pt\vrule}\hrule}}

\def\bse{\begin{eqnarray*}}
\def\ese{\end{eqnarray*}}
\def\be{\begin{eqnarray}}
\def\ee{\end{eqnarray}}
\def\bq{\begin{equation}}
\def\eq{\end{equation}}
\def\bse{\begin{eqnarray*}}
\def\ese{\end{eqnarray*}}

\begin{document}

\thispagestyle{empty} 
\baselineskip=28pt

\begin{center}
{\LARGE{\bf Conjugate Modeling Approaches for Small Area Estimation with Heteroscedastic Structure }}

\end{center}

\baselineskip=12pt

\vskip 2mm
\begin{center}
Paul A. Parker\footnote[2]{(\baselineskip=10pt to whom correspondence should be addressed)
Department of Statistics, University of California Santa Cruz,
1156 High St, Santa Cruz, CA 95064, paulparker@ucsc.edu}\footnote[1]{\label{CB}\baselineskip=10pt U.S. Census Bureau, 4600 Silver Hill Road, Washington, D.C. 20233-9100},
    Scott H. Holan\footnote[3]{\baselineskip=10pt Department of Statistics, University of Missouri
146 Middlebush Hall, Columbia, MO 65211-6100}\textsuperscript{\ref{CB}},  and
Ryan Janicki\textsuperscript{\ref{CB}}
\\
\end{center}
%
%
%
%
\vskip 4mm
\baselineskip=12pt 
\begin{center}
{\bf Abstract}
\end{center}

Small area estimation has become an important tool in official statistics, used to construct
estimates of population quantities for domains with small sample sizes. Typical area-level
models function as a type of heteroscedastic regression, where the variance for each domain is
assumed to be known and plugged in following a design-based estimate. Recent work has
considered hierarchical models for the variance, where the design-based estimates are used as
an additional data point to model the latent true variance in each domain. These hierarchical
models may incorporate covariate information, but can be difficult to sample from in
high-dimensional settings. Utilizing recent distribution theory, we explore a class of Bayesian
hierarchical models for small area estimation that smooth both the design-based estimate of the
mean and the variance. In addition, we develop a class of unit-level models for heteroscedastic
Gaussian response data. Importantly, we incorporate both covariate information as well as
spatial dependence, while retaining a conjugate model structure that allows for efficient
sampling. We illustrate our methodology through an empirical simulation study as well as an
application using data from the American Community Survey.

\baselineskip=12pt
\par\vfill\noindent
\par\medskip\noindent
\clearpage\pagebreak\newpage \pagenumbering{arabic}
\baselineskip=24pt

\section{Introduction}\label{sec: intro}

Population estimates for small areas or domains, {defined as geographic regions or
  demographic subpopulations,} based on survey data have garnered high demand in recent
years.  {These
  direct estimates, which use only domain-specific survey response data, may not be
  sufficiently precise for reliable inference due to small sample sizes and unreasonably high
  standard errors.}  To meet this demand, model based approaches, or small area estimation
(SAE) models, are commonly used in place of direct estimates. Area-level models such as the
popular Fay-Herriot (FH) model \citep{fay1979estimates} can incorporate covariates or other
dependencies in order to smooth the direct estimates { and improve precision by
  ``borrowing strength'' from areas with large sample size}. The FH model assumes that the
sample variance of the direct estimator is fixed and known, which is rarely the case in real
survey data settings. In practice, the sampling variance is estimated {from the
  survey data} and then plugged into the model. Sampling variances tend to vary across
geographic areas, thus, these area-level SAE models may be seen as a type of model for
heteroscedastic data.

Recently, a variety of extensions to the FH model have been considered in order to address the
issue of unknown sampling variances \citep{you2006small, you2021small}. For example, \citet{maiti2014prediction} use a hierarchical modeling
approach to jointly smooth both direct mean and variance
estimates. \citet{sugasawa2017bayesian} fit a similar model in a Bayesian setting. In both
cases, covariates may be used to aid in the variance smoothing, but in the Bayesian case,
\citet{sugasawa2017bayesian} required the use of a Metropolis-Hastings step within the Gibbs
sampler, due to a full-conditional distribution with unknown form.

Incorporating estimates of the sampling error variance of the direct estimator is a topic of
significant interest in the course of producing model-based estimates for official
statistics. In this direction, \citet{bradley2016bayesian} propose a model to include both the
direct estimates and an estimate of the sampling error variance in a Poisson area-level spatial
change-of-support model for the American Community Survey (ACS). The proposed model uses the
variance estimates as another data source and exploits the Poisson equidispersion assumption
(i.e., equal mean and variance) to condition on a common latent process.

The approach proposed here differs from \citet{bradley2016bayesian} and exploits the
distribution theory of \citet{bradley2019bayesian}. Specifically, we extend the approach
proposed in \citet{parker2021general} for heteroscedastic data to the SAE setting for both
area-level and unit-level models. Importantly, our approach yields models that are fully
conjugate for both the mean and variance regression parameters, and thus leads to extremely
computationally efficient estimation (see Sections~\ref{sec: methods} and \ref{sec:unit} for
additional detail).

Although SAE methods have a long and rich history, the literature on jointly modeling the mean
(direct estimates) and variance (sampling error variance) is significantly more recent by
comparison. For example, \citet{savitsky2022bayesian} propose a Bayesian nonparametric model
that jointly models the mean and the variance in the context of the Consumer Expenditure Survey
(CES). Similarly, \citet{polettini2017generalised} proposes a semiparametric Bayesian
Fay-Herriot model that shrinks both the means and variance. In general, these approaches can be
computationally expensive and, thus, there is a need for models that scale computationally to
meet the high-dimensional demands that are faced by statistical agencies and subject-matter
practitioners. The method proposed here meets this demand.

Outside of the SAE literature there exists a substantial literature on joint models for the
mean and covariance.  For example, \citet{pourahmadi1999joint, pourahmadi2011covariance} and
\citet{chen2003random}, among others, model Cholesky-based factorizations of unstructured
covariances. In general these methods are extremely useful; nevertheless, they are not
immediately applicable to the SAE problems considered here. In particular, the SAE setting is
comprised of a diagonal covariance (variance) structure with a model on the latent variances.
A comprehensive review of Cholesky-based joint mean and covariance modeling can be found in
\citet{pourahmadi2013high} and the references therein.

There has also been significant research on modeling covariance structure using covariates,
many of which arise in the spatio-temporal literature. For example, see
\citet{schmidt2011considering} and \cite{gladish2014covariate}, among others. The main
challenge that arises in this context is computational. In general, most of the proposed
methods proceed using Bayesian methods and lead to non-conjugate updates, necessitating
migration away from straightforward Gibbs sampling. One notable exception is proposed by
\citet{parker2021general}, which provides a starting point for the method proposed here.

 { This paper also proposes a novel unit-level SAE model for
  heteroscedastic Gaussian survey data.  The main complication of unit-level SAE modeling is
  accounting for the survey design in the model.  When the survey design is noninformative, in
  the sense that the distribution of the sampled response values is the same as the
  distribution of the unsampled values, the effect of the survey design can largely be
  ignored.  However, when the probability of selection in the survey is correlated with the
  response variable, the survey design is said to be informative, in which case the population
  distribution and the distribution for the sampled data will differ.  Under informative
  sampling, the survey design must be carefully accounted for in the SAE model to avoid biased
  estimates.}  To this end, we also propose a heteroscedastic unit-level model under
informative sampling using the pseudo-likelihood \citep{bin83, ski89, sav16}.  For a
comprehensive review of unit-level approaches to SAE see \citet{parker2019unit} and the
references therein.

Although the methodology proposed here is extremely general and applies to a broad set of applications that are encountered by data users and official statistical agencies, our motivating example considers income estimation and is related to the Small Area Income and Poverty Estimates Program (SAIPE). In particular, SAIPE produces income estimates for all U.S. states and counties \citep{bel16} and in many cases, the estimates may be used in the administration of federal programs and the allocation of federal funds to local jurisdictions.\footnote{\url{https://www.census.gov/programs-surveys/saipe/about.html}}. Therefore, using the American Community Survey (ACS) we demonstrate our proposed approach by estimating the average income by PUMA for the state of California.

This paper proceeds as follows. Section~\ref{sec: methods} provides background and introduces
our heteroscedastic model for area-level data. Unit-level models are presented in
Section~\ref{sec:unit}. To evaluate the effectiveness of our approach, an empirical simulation
study is provided in Section~\ref{sec: sim} and an analysis of ACS income
data is presented in Section~\ref{sec: DA}. Discussion is provided in Section~\ref{sec: disc}.

\section{Area-level Heteroscedastic Models}\label{sec: methods}

\subsection{Background}

The FH model \citep{fay1979estimates} is given by
\begin{align*}
    y_i | \theta_i,\sigma^2_i & \stackrel{ind}{\sim} \hbox{N}(\theta_i, \sigma^2_i), \;
      i=1,\ldots,d \\
    \theta_i &= \bm{x}_i'\bm{\beta} + \nu_i \\
    \nu_i & \stackrel{ind}{\sim} \hbox{N}(0, \sigma^2_{\nu}),
\end{align*} 
where $y_i$ is the direct estimate, $\sigma^2_i$ is the sampling variance of the direct
estimate {and is assumed known, and \(i = 1, \dots, d\) indexes the small areas of
  interest.}   Thus, the FH model may be seen as a
type of heteroscedastic data model where the variance is known.  Typically, the true value of
$\sigma^2_i$ is unknown, thus a design-based estimate, $s^2_i$ is plugged in instead.

In order to reflect the additional uncertainty attributed to $s_i^2,$ a second data model may be used. For example, \citet{you2006small} suggest the following model to address the issue of unknown sample variances,
\begin{align*}
    y_i | \theta_i,\sigma^2_i & \stackrel{ind}{\sim} \hbox{N}(\theta_i, \sigma^2_i), \; i=1,\ldots,d \\
    s_i^2 | \sigma_i^2 & \stackrel{ind}{\sim} \hbox{Gamma}\left(\frac{n_i-1}{2}, \frac{n_i-1}{2\sigma^2_i} \right), \; i=1,\ldots,d \\
    \theta_i &= \bm{x}_i'\bm{\beta} + \nu_i \\
    \nu_i & \stackrel{ind}{\sim} \hbox{N}(0, \sigma^2_{\nu}) \\
    \sigma^2_i & \stackrel{ind}{\sim} \hbox{IG}(a_i,b_i),
\end{align*} 
{where $n_i$ represents the sample size in area $i$ and \(\hbox{IG}(a, b)\) denotes
  an inverse gamma distribution with shape parameter \(a\) and scale parameter \(b\).}  In
principal, the data model for $s_i^2$ given here is only valid in the case of a simple random
sample within area $i$. For complex sample designs, it may be important to give careful consideration
to the degrees of freedom. Although estimation of the appropriate degrees of freedom is beyond the scope of this work, for more discussion on this matter, see \citet{maples2009small}.

\citet{sugasawa2017bayesian} provide a Bayesian extension of this approach that considers
covariates in the variance model by letting $\sigma^2_i \stackrel{ind}{\sim}
\hbox{IG}\left(a_i,b_i \hbox{exp}(\bm{x}_{2i}'\bm{\beta}_2)\right)$. Although the use of
covariates here may improve small area estimates, \citet{sugasawa2017bayesian} required the use
of a Metropolis-Hastings sampler for $\bm{\beta}_2$, which can be {extremely} difficult to tune,
especially in high dimensions.

\subsection{Conjugate Priors for Heteroscedastic Models}

The foundation of our modeling framework is the multivariate log-Gamma distribution (MLG), introduced by \cite{bradley2018computationally} and \cite{bradley2019bayesian}. The MLG distribution was initially developed in order to model dependent data using a Poisson likelihood. The density for the MLG distribution is given as 
\begin{equation}
    f(\bm{y})=\hbox{det}(\bm{V}^{-1})
    \left\{ \prod_{i=1}^n \frac{\kappa_i^{\alpha_i}}{\Gamma(\alpha_i)}\right\}
    \hbox{exp}\left[\bm{\alpha}' \bm{V}^{-1}(\bm{y- \mu}) -
    \bm{\kappa}' \hbox{exp}\left\{\bm{V}^{-1}(\bm{y- \mu}) \right\}\right],
\end{equation} denoted by $\hbox{MLG}(\bm{\mu}, \mathbf{V}, \bm{\alpha}, \bm{\kappa})$.  The length $n$ vector $\bm{\mu}$ acts as a centrality parameter and the $n \times n$ matrix $\bm{V}$ controls the correlation structure. The  length $n$ vectors $\bm{\alpha}$ and $\bm{\kappa}$ are shape and rate parameters respectively.  Sampling from $\bm{Y}\sim \hbox{MLG}(\bm{\mu}, \mathbf{V}, \bm{\alpha}, \bm{\kappa})$ is straightforward using the following steps:
\begin{enumerate}
        \item Generate a vector $\mathbf{g}$ as $n$ independent Gamma random variables with shape $\alpha_i$ and rate $\kappa_i$, for $i=1,\ldots,n$
        \item Let $\mathbf{g}^*=\hbox{log}(\mathbf{g})$
        \item Let $\mathbf{Y}=\mathbf{V g}^* + \bm{\mu}$.
    \end{enumerate} 

In most cases, Bayesian inference using the MLG prior distribution also requires simulation from the conditional multivariate log-Gamma distribution (cMLG). Letting $\mathbf{Y} \sim \hbox{MLG}(\bm{\mu},  \mathbf{V}, \bm{\alpha}, \bm{\kappa})$, \cite{bradley2018computationally} show that $\mathbf{Y}$ can be partitioned into $(\mathbf{Y_1}', \mathbf{Y_2}')'$, where $\mathbf{Y_1}$ is $r$-dimensional and $\mathbf{Y_2}$ is $(n-r)$-dimensional. The matrix $\mathbf{V}^{-1}$ is also partitioned into $\left[\mathbf{H \; B}  \right]$, where $\mathbf{H}$ is an $n \times r$ matrix and $\mathbf{B}$ is an $ n \times (n - r)$ matrix. Then 
$$\bm{Y_1} | \bm{Y_2} = \bm{d}, \bm{\mu}^*, \bm{H}, \bm{\alpha}, \bm{\kappa} \sim \hbox{cMLG}(\bm{\mu}^*, \bm{H}, \bm{\alpha}, \bm{\kappa})$$ with density
\begin{equation}
    M \hbox{exp} \left\{\bm{\alpha}' \bm{H Y_1} 
     - \bm{\kappa}' \hbox{exp}(\bm{H Y_1} - \bm{\mu}^*)\right\} I\left\{(\bm{Y_1}' , \bm{d}')' \in \mathcal{M}^n \right\},
\end{equation} where $\bm{\mu}^*=\mathbf{V}^{-1}\bm{\mu} - \mathbf{Bd}$, and $\mathit{M}$ is a normalizing constant.  \cite{bradley2018computationally} show that it is also straightforward to sample from the cMLG distribution using $(\mathbf{H}'\mathbf{H})^{-1}\mathbf{H}'\mathbf{Y}$, where $\mathbf{Y}$ is sampled from $\hbox{MLG}(\bm{\mu}, \mathbf{I}, \bm{\alpha}, \bm{\kappa})$.

Another key result given by \cite{bradley2018computationally} is that $\hbox{MLG}(\mathbf{c}, \alpha^{1/2}\mathbf{V}, \alpha \mathbf{1}, \alpha \mathbf{1})$ converges in distribution to a multivariate normal distribution with mean $\mathbf{c}$ and covariance matrix $\mathbf{VV'}$ as the value of $\alpha$ approaches infinity. This allows for the computational benefits of MLG priors while maintaining effectively the same shape and structure as a Gaussian prior.

Although the original use of the MLG distribution was as a prior in high-dimensional Poisson regression, recently \cite{parker2021general} found that the MLG distribution acts as a conjugate prior for variance regression when using a negative log link function. This insight is the basis for our proposed approach.

\subsection{Proposed Area-Level Approach}

In order to account for the additional uncertainty around $s_i^2$ in {the context of} area-level SAE, we construct a joint model for the direct point estimate and variance. Critically, this model relies on the MLG distribution as a conjugate prior for the variance regression parameters.  Our heteroscedastic area-level model (HALM) is given as

    \begin{align*}
    {y_i} | \theta_i,\sigma^2_i & \stackrel{ind}{\sim} \hbox{N}(\theta_i, \sigma^2_i), \; i=1,\ldots,d \\
    {s_i^2} | \sigma_i^2 & \stackrel{ind}{\sim} \hbox{Gamma}\left(\frac{{n_i}-1}{2}, \frac{{n_i}-1}{2\sigma^2_i} \right), \; i=1,\ldots,d \\
    \theta_i &= {\bm{x}_i}'\bm{\beta}_1 +  \eta_{1i} \\
    -\hbox{log}(\sigma^2_i) &= {\bm{x}_i}'\bm{\beta}_2 +  \eta_{2i} \\
    \bm{\eta}_1 | \sigma^2_{\eta_1} & \sim \hbox{N}(\bm{0}, \sigma^2_{\eta_1} \bm{\hbox{I}}) \\
    \bm{\eta}_2 | \sigma^2_{\eta2} & \sim \hbox{MLG}(\bm{0}, \alpha^{1/2} \sigma_{\eta_2} \bm{\hbox{I}}, \alpha \bm{1}, \alpha \bm{1}) \\
    \bm{\beta}_1 & \sim \hbox{N}(\bm{0}, \sigma^2_{\beta_1} \bm{\hbox{I}}) \\
        \bm{\beta}_2  & \sim \hbox{MLG}(\bm{0}, \alpha^{1/2} \sigma_{\beta_2} \bm{\hbox{I}}, \alpha \bm{1}, \alpha \bm{1}) \\
        \sigma^2_{\eta_1} & \sim \hbox{IG}(a,b) \\
        \sigma_{\eta_2} & \sim \hbox{N}^+(0,c).
\end{align*} 
Here, $y_i$ represents the direct estimate of an unknown population quantity, $\theta_i,$ while
$s_i^2$ represents the design-based variance around this estimate. The unknown population
quantity is written as a linear combination of the length $p$ vector of covariates, $\bm{x}_i,$
as well as an area level random effect, $\eta_{1i}.$ The unknown variance, $\sigma_i^2,$ is
modeled using the negative log link function as a linear combination of $\bm{x}_i$ as well as
an additional area-level random effect, $\eta_{2i}.$ In order to establish conjugate
full-conditional distributions, $\beta_2$ takes on a MLG prior distribution that is
asymptotically equivalent to a $\hbox{N}(\bm{0}, \sigma^2_{\beta} \bm{\hbox{I}})$
distribution. Similarly, conditional on $\sigma^2_{\eta_2},$ $\bm{\eta}_2$ takes a MLG prior
distribution that is asymptotically equivalent to a $\hbox{N}(\bm{0},
\sigma^2_{\eta_2}\bm{\hbox{I}})$ distribution. Finally, we place a conjugate inverse Gamma
prior distribution on $\sigma^2_{\eta_1}$ as well as a half-normal prior on $\sigma_{\eta_2}.$
We note that this prior for $\sigma_{\eta_2}$ is not conjugate, and thus requires a
Metropolis-Hastings step. However, this is only for a single parameter and we have found that there is very
little effect on the mixing of the MCMC due to this. We use a random-walk Metropolis-Hastings step with a Normal distribution truncated below at zero for a proposal distribution, although, depending on the setting, it may be helpful to consider other proposals. The model is completed by specification of
$\alpha, \sigma^2_{\beta_1}, \sigma^2_{\beta_2}, a, b, c >0.$ In practice, we work with relatively diffuse priors by using
$\sigma^2_{\beta_1}=\sigma^2_{\beta_2}=1000,$ $a=b=0.5,$ and $c=5.$ The value for $\alpha$ should be sufficiently
large to invoke the asymptotic equivalence to the multivariate normal distribution. Similar to
\cite{bradley2018computationally}, we have found $\alpha=1000$ to be sufficient for our
purposes, although in some other cases this may be data dependent.

Often within SAE, more precise estimates can be generated through the use of spatial dependence modeling (e.g., see \cite{marhuenda2013small,porter2015small}). This motivates the need for spatially correlated prior structures for the mean and variance models. To this end, we develop a spatial variant of the HALM model, termed the spatial heteroscedastic area-level model (SHALM). This model is similar to HALM, with the replacement of the prior structure for $\bm{\eta}_1$ and $\bm{\eta}_2,$
\begin{align*}
        \bm{\eta}_1 | \sigma^2_{\eta_1} & \sim \hbox{N}\left(\bm{0}, \sigma^2_{\eta_1}(\bm{D} - \bm{W})^{-1}\right) \\
    \bm{\eta}_2 | \sigma^2_{\eta2} & \sim \hbox{MLG}(\bm{0}, \alpha^{1/2} \sigma_{\eta_2}(\bm{D} - \bm{W})^{-1/2} , \alpha \bm{1}, \alpha \bm{1}).
\end{align*}
Here, the $d \times d$ matrix $\bm{W}$ is an area-level adjacency matrix, with entry $W_{ij}=1$ if areas $i$ and $j$ share a border and $W_{ij}=0$ otherwise. By convention, an area is not considered a neighbor of itself, resulting in a zero value for all diagonal elements. The $d \times d$ matrix $\bm{D}$ is a diagonal matrix, where the $i$th entry corresponds to the number of neighbors shared by area $i,$ or equivalently, the sum of the $i$th row of the matrix $\bm{W}.$ This prior for $\bm{\eta}_1$ is known as the intrinsic conditional autoregressive (ICAR) prior \citep{besag1991bayesian}. Similarly, the prior for $\bm{\eta}_2$ is asymptotically equivalent to an ICAR prior.

\section{Unit-level Volatility Models}\label{sec:unit}

An increasingly common alternative to area-level models for SAE is that of unit-level modeling. Unit-level models opt to model the survey data directly rather than the design-based estimates as in the area-level case. For example, the basic unit-level model (BULM) was introduced by \cite{bat88} and is written as
    \begin{align*}
        {y_{ij}} & \stackrel{ind}{\sim} \hbox{N}(\mu_{ij}, \sigma^2), \; j \in \mathcal{S} \\
        \mu_{ij} &= {\bm{x}_{ij}}' \bm{\beta} + \eta_i \\
        \eta_i & \stackrel{iid}{\sim} \hbox{N}(0, \sigma^2_{\eta}).
    \end{align*} Here, $y_{ij}$ is the response for unit $j$ in the sample, $\mathcal{S}$, residing in area $i$, while $\bm{x}_{ij}$ is a vector of unit-level covariates. The area-level random effects, $\eta_i,$ allow for dependence among respondents within the same area. For this model, as well as other unit-level modeling approaches, the model may be fit using the observed sample data and predictions can be made for the entire population. In essence, this results in a synthetic population that may be aggregated as necessary to produce area-level estimates at the desired spatial resolution.

One major limitation of the BULM is that it assumes the survey design to be ignorable. Many
surveys result in an informative sampling scheme in which there is a relationship between the
response of interest and the unit probabilities of selection.  {Let
  \(\mathcal{U}_i\) be an enumeration of the finite population in area \(i\), and let
  \(\mathcal{S}_i \subset \mathcal{U}_i\) be the survey sample from area \(i\) selected
  according to a known sampling scheme with inclusion probabilities \(P \left( j \in
  \mathcal{S}_i \right) = \pi_{ij}\).  Define the survey weights as \(w_{ij} = 1 / \pi_{ij}\).
  Informative survey designs occur when survey inclusion indicators are correlated with survey
  response variables, even after conditioning on observable covariates and design
  variables.} In these
situations, use of a model that does not consider the survey design may result in large
biases. One popular solution to this problem is the use of an exponentially weighted
pseudo-likelihood \citep{bin83, ski89}. More recently, \cite{sav16} popularized the use of a
pseudo-likelihood in general Bayesian model settings. This results in a pseudo-posterior
distribution that is proportional to the product of the pseudo-likelihood and the prior
distribution,
$$
\widehat{\pi}(\bm{\theta} | \mathbf{y}, \mathbf{\tilde{w}}) \propto \left\{ \prod_{j \in \mathcal{S}} f(y_{ij} | \bm{\theta})^{\tilde{w}_{ij}} \right\} \pi (\bm{\theta}).
$$ In this case, $\tilde{w}_{ij}$ represents the survey weights after scaling to sum to the
sample size. For example, a Bayesian pseudo-likelihood alternative to the BULM may be written
as
    \begin{equation}\label{EQ: PLBULM}
    \begin{aligned}
        \bm{y} | \bm{\mu}, \sigma^2 & \propto \prod_{j \in \mathcal{S}}\hbox{N}(y_{ij}|\mu_{ij}, \sigma^2)^{\tilde{w}_{ij}} \\
        \mu_{ij} &= {\bm{x}_{ij}}' \bm{\beta} + \eta_i \\
        \eta_i & \stackrel{iid}{\sim} \hbox{N}(0, \sigma^2_{\eta}).
        \end{aligned}
    \end{equation} Although there are many other approaches to account for an informative sample design, our focus here will be strictly on the Bayesian pseudo-likelihood. For an overview of alternative approaches, see \cite{parker2019unit}.
    
Another limitation of the BULM is the assumption of constant variance across survey units. In practice, the dispersion of a particular response of interest may vary along with certain covariates or by geographic region.

\subsection{Proposed Unit-Level Approach}

In order to address limitations of the BULM, we propose a unit-level model that accounts for a
possibly informative sample design while also relaxing the constant variance assumption. This
approach is termed the heteroscedastic unit-level model (HULM) and is written
as

    \begin{align*}
        \bm{y} | \bm{\mu}, \bm{\sigma^2} & \propto \prod_{j \in \mathcal{S}} \hbox{N}({y_{ij}} | \mu_{ij}, \sigma^2_{ij})^{\tilde{w_{ij}}} \\
        \mu_{ij} &= {\bm{x}_{ij}}'\bm{\beta}_1 + \eta_{1i} \\
        -\log(\sigma^2_{ij}) &= {\bm{x}_{ij}}'\bm{\beta}_2 + \eta_{2i} \\
        \bm{\eta}_1 | \sigma^2_{\eta_1} & \sim \hbox{N}(\bm{0}, \sigma^2_{\eta_1}) \\
    \bm{\eta}_2 | \sigma^2_{\eta2} & \sim \hbox{MLG}(\bm{0}, \alpha^{1/2} \sigma_{\eta_2} \bm{\hbox{I}}, \alpha \bm{1}, \alpha \bm{1}) \\
    \bm{\beta}_1 & \sim \hbox{N}(\bm{0}, \sigma^2_{\beta_1}) \\
        \bm{\beta}_2  & \sim \hbox{MLG}(\bm{0}, \alpha^{1/2} \sigma_{\beta_2} \bm{\hbox{I}}, \alpha \bm{1}, \alpha \bm{1}) \\
                \sigma^2_{\eta_1} & \sim \hbox{IG}(a,b) \\
        \sigma_{\eta_2} & \sim \hbox{N}^+(0,c).
    \end{align*} 
This approach uses a Gaussian pseudo-likelihood with individual mean and variance to model the
data. The individual means, $\mu_{ij}$ are written as a linear combination of the length $p$
covariate vector, $\bm{x}_{ij}$, as well as an area-level random effect, $\eta_{1i}.$ Using a
negative log link function, the individual variances, $\sigma^2_{ij}$ are also modeled as a
linear combination of $\bm{x}_{ij}$ and an additional area-level random effect, $\eta_{2i}.$
    
The model structure for $\bm{\eta}_1,$ $\bm{\eta}_2,$ $\bm{\beta}_1,$ and $\bm{\beta}_2$ is
identical to the HALM. In particular, $\bm{\eta}_1$ and $\bm{\eta}_2$ are modeled
hierarchically with unknown variance parameters while both $\bm{\eta}_2$ and $\bm{\beta}_2$
take MLG prior distributions to allow for computational feasibility. Although not directly
explored here, it is straightforward to extend this approach to consider spatially correlated
random effects similar to the SHALM; for example, see \cite{sun2022analysis}.

\section{Empirical Simulation Study}\label{sec: sim}

In many cases, the decision whether to use an area-level vs. a unit-level model will depend on whether an analyst has access to unit-level microdata. Additionally, in situations where many areas contain little or no data, an area-level approach may be infeasible due to the lack of appropriate design-based estimates. Here, we aim to compare a variety of both area-level and unit-level models by devising an appropriate simulation study. However, we note that in practice, it is possible that some subset of the models explored here may not be appropriate.

To construct our simulation study, we require a population of individuals that we may sample from. Rather than using a synthetic population drawn from some parametric distribution, we instead take an existing survey dataset and treat it as our population in order to preserve many of the characteristics associated with the real data. In particular, we use the public-use microdata sample (PUMS) from the American Community Survey (ACS). We restrict our scope to the 1-year PUMS data for the state of California only. This empirical population contains roughly 179,000 individuals. Each individual is associated with a geographic region known as the public-use microdata area (PUMA). The state of California contains 265 PUMAs. Ultimately our goal is to estimate average income by PUMA using a sample from this population.

We consider two different approaches for subsampling from the empirical population. First, we take a stratified random sample by PUMA with a simple random sample without replacement of 5 observations per PUMA. Second, we take a probability proportional to size sample using the Poisson method \citep{brewer1984poisson} with an expected total sample size of 1,000. For the second sample design, we construct the size variables as $\exp\left(2 + 0.3 \times w_{ij} + 0.3 \times \tilde{y}_{ij}\right),$ where $w_{ij}$ is the original scaled survey weight and $\tilde{y}_{ij}$ is the scaled income for unit $j$ in area $i$. The use of income in the size variable enforces an informative design. For both sampling approaches, we repeat the sampling and estimation procedure $K=100$ times. Horvitz-Thompson direct estimates of mean income are calculated for each PUMA.

We consider two different unit-level models for this study. First, we present the  Bayesian pseudo-likelihood alternative of the BULM (PL-BULM) given in (\ref{EQ: PLBULM}). We also compare to the proposed HULM. We note that exploratory analysis indicated that the spread of income did not vary by PUMA, so for the HULM, we constrain $\bm{\eta}_2 = \bm{0}.$ For both unit-level approaches, we model income after taking a log transformation, and we use age, sex, and race as covariates. We note that at the unit level, Gaussian models for income are a starting point, but further work is necessary in this area. For the PUMS data in particular, there is some rounding and top-coding that occurs as a disclosure avoidance mechanism. The unit-level models explored here are adequate for characterization of the first moment of income, but a more complex model seems necessary to adequately model the full distributional uncertainty. 

We also consider four area-level models. First, we fit the basic Fay-Herriot model (FH). Second, we consider the model used by \cite{sugasawa2017bayesian} that shrinks both the design-based mean and variance estimates (STK). Lastly, we fit both the proposed HALM and SHALM. All area-level models are fit after log transforming the design-based estimates of income and using the delta method for variance estimates. Log population size was used as a covariate. All models were fit using MCMC with 3,000 iterations, discarding the first 1,000 iterations as burn-in. Convergence was assessed via traceplots of the sample Markov chains, with no lack of convergence detected.

We are primarily interested in two forms of assessment for these models. First, we examine the root mean squared error (RMSE) of our point estimates,
    $$\sqrt{\sum_{k=1}^K \frac{(\widehat{\theta}_k - \theta)^2}{K}}. $$ Here, $\theta$ represents the true population quantity of interest while $\widehat{\theta}_k$ represents an estimate for sample dataset $k$.
 RMSE has the desirable property of being composed of both a bias and variance term. We also consider the interval score \citep{gneiting2007strictly} for our 95\% credible interval estimates,
     $$ {
    \frac{1}{K} \sum_{k=1}^K \left\{(u_k - \ell_k) + \frac{2}{\alpha}(\ell_k - \theta)I(\theta < \ell_k) + \frac{2}{\alpha}(\theta - u_k)I(\theta > u_k)\right\}},
    $$ where $\alpha = 0.05,$ $u_k$ is the upper bound of the interval, and $\ell_k$ is the lower bound of the interval for sample dataset $k.$ For the interval score, a lower score is desirable. Thus, narrow intervals are rewarded, but a penalty is incurred if the interval misses the true value.
    
    Results for the stratified sampling design and the probability proportional to size design are summarized in Tables \ref{tab1} and \ref{tab2} respectively. All results are averaged across PUMAs. RMSE is presented relative to the direct estimator, where a value less than one indicates a reduction in RMSE relative to the direct estimator. For the stratified sampling design, all models were able to reduce the RMSE relative to the direct estimator with the exception of the PL-BULM. Thus, the proposed HULM was able to offer substantial improvement over a model that assumes constant variance. For the HULM, the interval estimates were also improved relative to the PL-BULM, although as discussed previously, further model development here is desirable. In terms of area-level approaches, the FH model performed worst in terms of both RMSE and interval score. Therefore, shrinkage of both the mean and variance appears to be important in order to improve the quality of generated estimates. The HALM and STK models resulted in quite similar RMSE and interval scores. Finally, the SHALM was able to leverage spatial dependence resulting in the lowest RMSE and interval scores across all models. Results were similar for the probability proportional to size design, indicating robustness to the assumption of a simple random sample within each area used in the variance shrinkage model.

    \begin{table}[H]
\centering
\resizebox{\textwidth}{!}{\begin{tabular}{lrrrr}
  \hline
Estimator & Rel. RMSE  & Abs. Bias ($\times 10^4$) & Cov. Rate & Int. Score ($\times 10^4$) \\ 
  \hline
PL-BULM & 1.080 & 20.299 & 0.368 & 30.570 \\ 
  HULM & 0.687 & 11.356 & 0.596 & 14.720 \\ 
  \hdashline
  FH & 0.694 & 5.126 & 0.894 & 8.790 \\ 
  HALM & 0.640 & 7.677 & 0.956 & 6.695 \\ 
  SHALM & \textbf{0.561} & 6.759 & 0.933 & \textbf{6.031} \\ 
  STK & 0.636 & 6.958 & 0.952 & 6.648 \\ 
   \hline
\end{tabular}}
\caption{Empirical simulation results for stratified random sampling by Public-Use Microdata Area (PUMA) using the 2018 1-year American Community Survey public-use microdata sample. All results are averaged across PUMAs. RMSE is presented relative to the direct estimator.} 
\label{tab1}
\end{table}

\begin{table}[H]
\centering
\resizebox{\textwidth}{!}{\begin{tabular}{lrrrr}
  \hline
Estimator & Rel. RMSE  & Abs. Bias ($\times 10^4$) & Cov. Rate & Int. Score ($\times 10^4$) \\ 
  \hline
PL-BULM & 0.766 & 19.607 & 0.392 & 30.846 \\ 
  HULM & 0.646 & 16.509 & 0.461 & 23.574 \\ 
  \hdashline
  FH & 0.492 & 6.815 & 0.955 & 8.855 \\ 
  HALM & 0.442 & 9.765 & 0.958 & 6.800 \\ 
  SHALM & \textbf{0.401} & 8.481 & 0.913 & \textbf{6.267} \\ 
  STK & 0.429 & 9.195 & 0.958 & 6.468 \\ 
   \hline
\end{tabular}}
\caption{Empirical simulation results for probability proportional to size sampling using the 2018 1-year American Community Survey public-use microdata sample. All results are averaged across PUMAs. RMSE is presented relative to the direct estimator.} 
\label{tab2}
\end{table}

\begin{figure}[H]
\includegraphics[width=\textwidth]{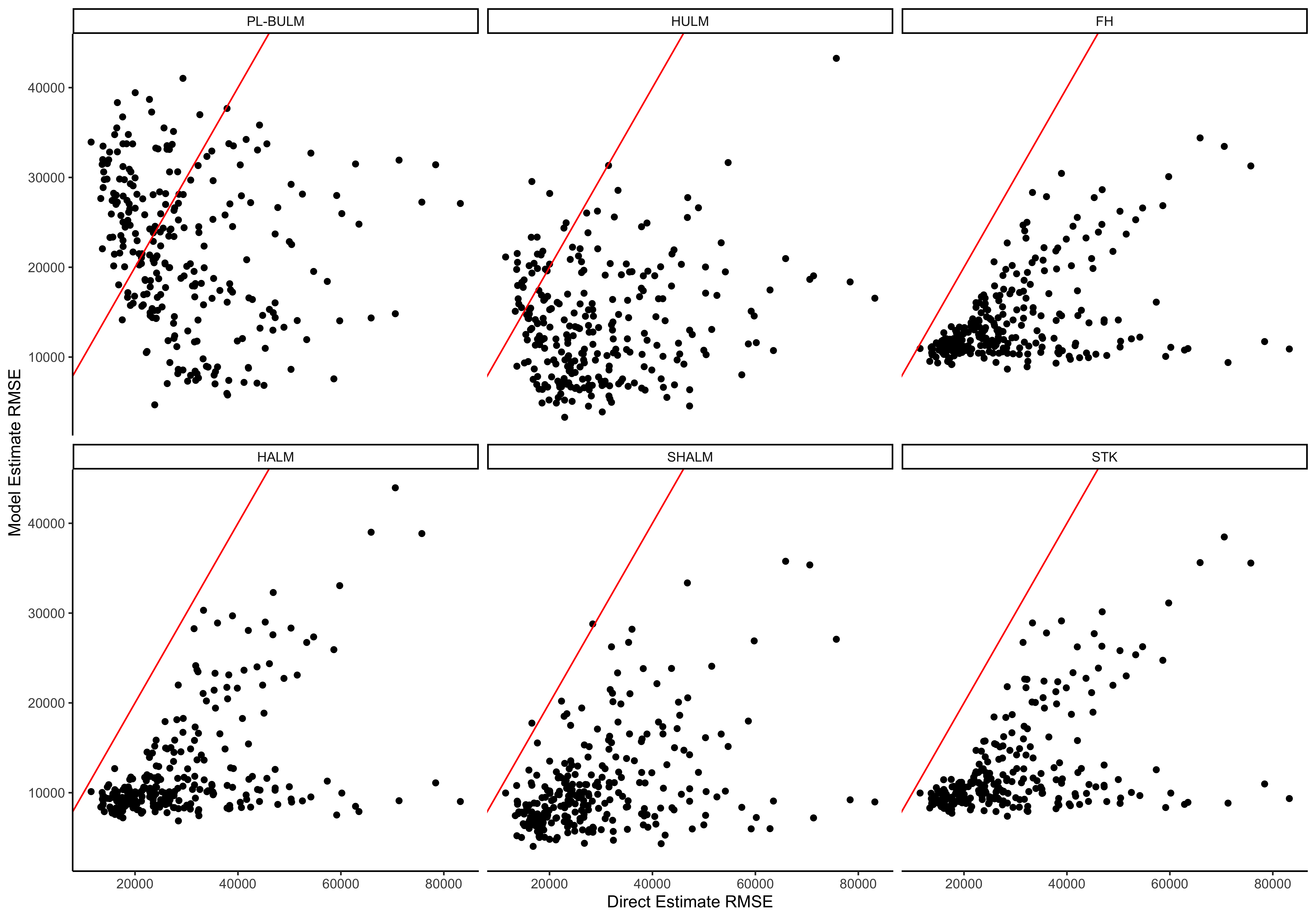}
\caption{Empirical simulation results of Direct vs. Model-based RMSE by Public-Use Microdata Area (PUMA) for stratified random sampling using the 2018 1-year American Community Survey public-use microdata sample.} 
\label{FIG: Strat}
\end{figure}

\begin{figure}[H]
\includegraphics[width=\textwidth]{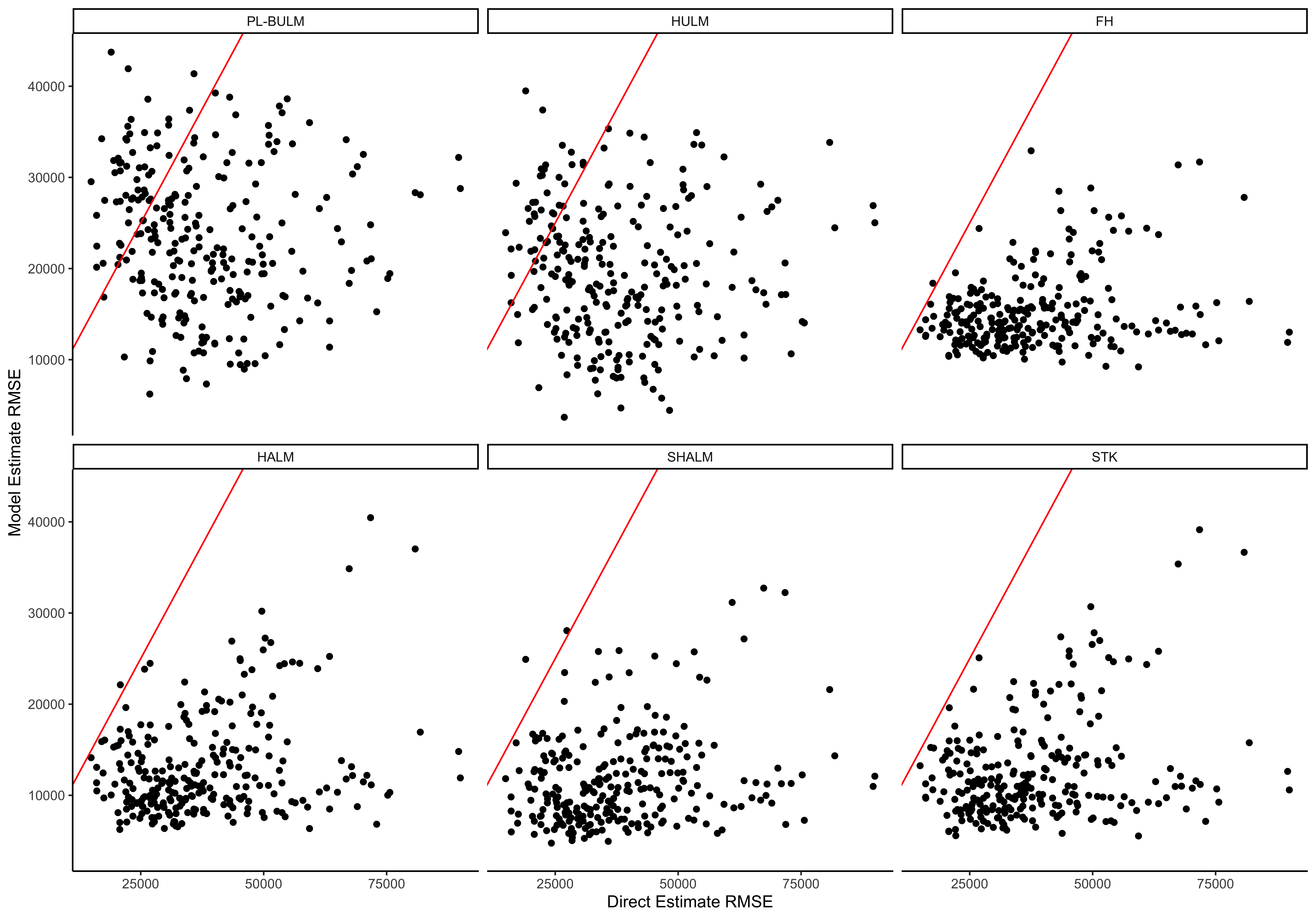}
\caption{Empirical simulation results of Direct vs. Model-based RMSE by Public-Use Microdata Area (PUMA) for probability proportional to size sampling using the 2018 1-year American Community Survey public-use microdata sample.} 
\label{FIG: PPS}
\end{figure}

We also compare the RMSE by PUMA between the direct estimates and each model based estimate. Figure \ref{FIG: Strat} presents these results for the stratified sample design. With the exception of the PL-BULM, most to all PUMAs experience a reduction in RMSE relative to the direct estimator, as indicated by points that fall below the one-to-one line. Among the area-level models, there is a general downward shift in points for the STK, HALM, and STK models compared to the FH model. This indicates a general reduction in RMSE for most areas when compared to the FH model. Similarly, the SHALM appears to have a general downward shift when compared to the STK and HALM methods. Similar results are presented for the probability proportional to size design in Figure \ref{FIG: PPS}, for which similar patterns hold.

Taken collectively, this simulation illustrates that heteroscedastic modeling techniques may be used to improve the quality of small area estimates. At the area level, these techniques may be used to simultaneously shrink both the design-based means as well as variances. At the unit-level, our framework allows for respondent specific variances when modeling continuous data. In both cases, our framework has the potential to improve the precision of associated small area estimates relative to approaches that do not have flexible models for the variance. 

\section{Analysis of American Community Survey Income Data}\label{sec: DA}

One important application of small area estimation techniques is estimation of mean or median income for various geographies. For example, the Small Area Income and Poverty Estimates Program (SAIPE) produces income estimates for all U.S. states and counties \citep{bel16}. In many cases, the estimates produced by SAIPE or similar programs may be used to allocate critical federal aid. Thus, improving the quality of model-based estimates for various outcomes such as income constitutes an important research problem. To this end, we demonstrate an application of our proposed SHALM approach by estimating the average income by PUMA for the state of California using the 2018 1-year ACS PUMS sample. The SHALM is fit analogously to Section \ref{sec: sim}, with the exception that the entire PUMS dataset was used rather than a subsample.

\begin{figure}[H]
\includegraphics[width=\textwidth]{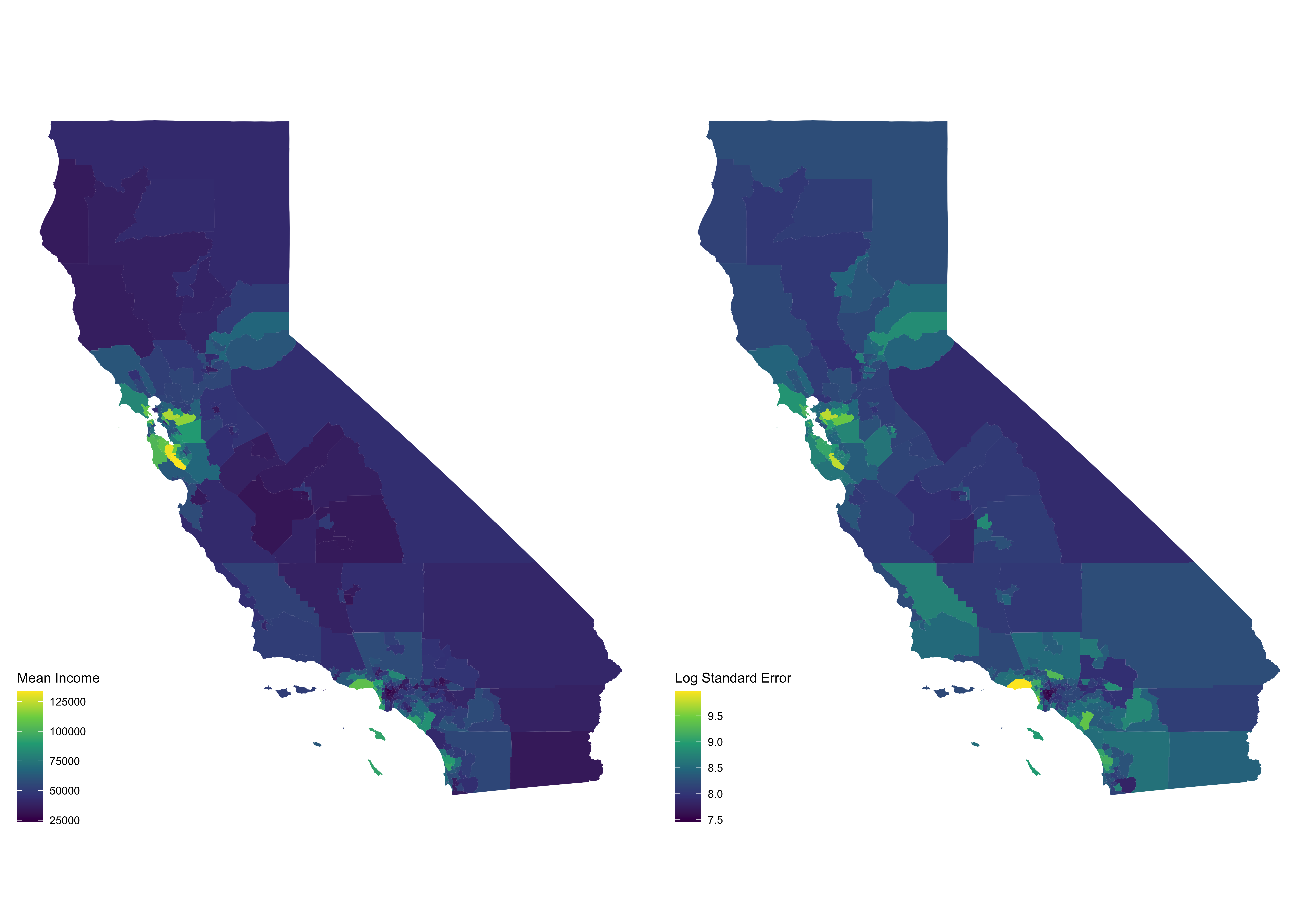}
\caption{Model-based estimates of mean income by public use microdata area along with associated log transformed standard errors. Estimates are constructed using the 2018 1-year American Community Survey Public Use Microdata Sample.} 
\label{FIG: Ests}
\end{figure}

The model-based estimates of mean income by PUMA are shown in Figure \ref{FIG: Ests}, along with their standard errors. The estimates are generally as expected with higher incomes around city centers with high cost of living, such as the San Francisco Bay area and Los Angeles, and lower incomes in more rural parts of the state. Uncertainty is higher in areas that had lower sample sizes, but also in some areas with very high estimated income. For these counties, there was considerably more spread in the observed incomes, which contributes to the uncertainty around the estimated mean.

\begin{figure}[H]
\includegraphics[width=\textwidth]{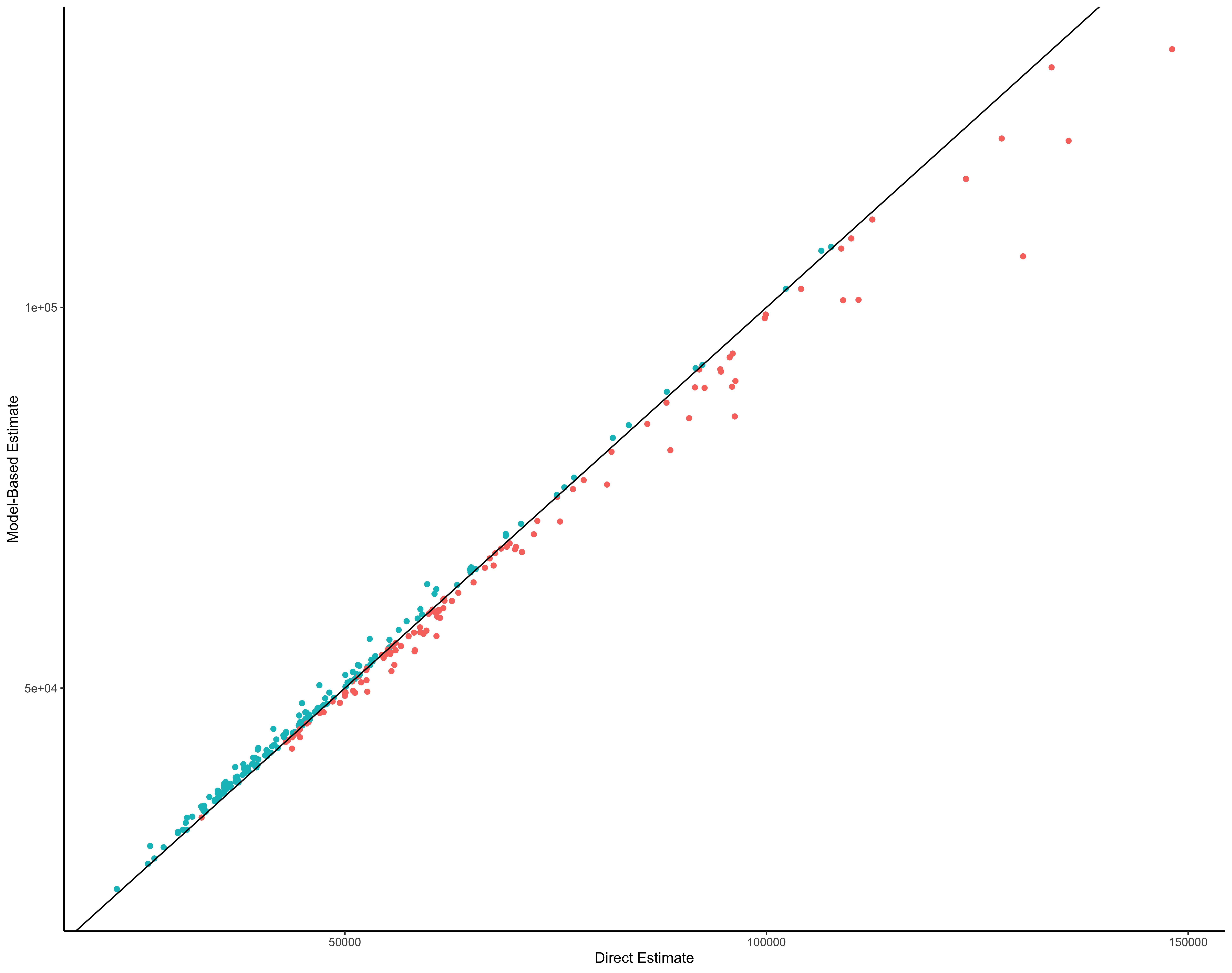}
\caption{Direct vs. Model-based estimates of mean income by public use microdata area. Estimates are constructed using the 2018 1-year American Community Survey Public Use Microdata Sample.} 
\label{FIG: Compare}
\end{figure}

We also compare the model-based estimates to the direct estimates in Figure \ref{FIG: Compare}. Here, we see that the two estimates generally agree, with points falling close to the one-to-one line. However, we also see that the model-based approach tends to result in slightly higher estimates for areas with low average income and slightly lower estimates for areas that exhibited high average income.

\section{Discussion}\label{sec: disc}
Small area/domain estimation is an area of wide-spread interest both for data users and official statistical agencies. Consequently, there has been significant research on both area-level and unit-level models with the goal of improving the precision of target estimates. Nevertheless, model-based approaches typically focus on modeling the mean, with a few notable exceptions. As illustrated here, in the presence of heteroscedasticity, simultaneously modeling both the mean and variance can achieve estimates with both reduced MSE and superior frequentist coverage properties.

By extending \citet{bradley2019bayesian} and \citet{parker2021general}, our approach to simultaneously modeling the mean and variance produces fully conjugate updates for hard-to-estimate parameters and is, therefore, extremely computationally efficient. Importantly, our approach is extremely flexible and allows for the incorporation of  spatial dependence and covariates in the portion of the model for the variance.

Although our main focus is on area-level modeling, we also introduce unit-level models that account for informative sampling through the use of the pseudo-likelihood. These models can be extremely effective in situations where tabulations are desired for custom-geographies and/or situations when internal aggregation consistency is desired.

Our motivating example focused on modeling income and showcased the gains achieved from using our proposed approach. Nevertheless, for this example, our area-level models outperformed the models that were estimated at the unit-level. One reason for this is that the Gaussian assumption at the unit-level may not be an optimal starting point for this application due to data issues that arise from disclosure avoidance mechanisms. In this direction, there are several avenues for future work, including extensions to non-Gaussian data or Gaussian mixtures for both unit- and area-level models. In addition, future work also includes the extension to multivariate applications and applications where data integration is advantageous.

\section*{Acknowledgements}
 This research was partially supported by the U.S.~National Science Foundation (NSF) under NSF grant SES-1853096. 
This article is released to inform interested parties of ongoing research and to encourage discussion. The views expressed on statistical issues are those of the authors and not those of the NSF or U.S. Census Bureau.

\bibliographystyle{jasa}

\end{document}